\documentstyle[12pt]{article}
\renewcommand{\theequation}
{\mbox{\arabic{section}.\arabic{equation}}}
\newtheorem{theorem}{Theorem}[section]

\newcommand{\beq}{\begin{equation}}
\newcommand{\eeq}{\end{equation}}
\newcommand{\bea}{\begin{eqnarray}}
\newcommand{\eea}{\end{eqnarray}}
\newcommand{\D}{\displaystyle}

\newcommand{\ssc}{\scriptscriptstyle}
\newcommand{\U}{\hat{U}}
\newcommand{\hnu}{\hat{\nu}}
\newcommand{\hmu}{\hat{\mu}}
\newcommand{\tr}{{\rm tr}}
\newcommand{\Z}{{\cal Z}}
\newcommand{\vev}[1]{\Big\langle #1 \Big\rangle}
\input epsf
\begin{document}

\hfill \vbox{\hbox{UCLA/98/TEP/14}
             \hbox{hep-lat/9806030}} 
\begin{center}{\Large\bf Absence of confinement in the absence 
of vortices}\\[2cm] 
{\bf Tam\'as G. Kov\'acs}\footnote{Research supported by 
DOE grant DE-FG02-92ER-40672} \\
{\em Department of Physics, University of Colorado, Boulder 
CO 80309-390}\\
{\sf e-mail: kovacs@eotvos.Colorado.EDU}\\[5mm] 

and\\[5mm]

{\bf E. T. Tomboulis}\footnote{Research supported by 
NSF grant NSF-PHY 9531023}\\
{\em Department of Physics, UCLA, Los Angeles, 
CA 90095-1547}\\
{\sf e-mail: tombouli@physics.ucla.edu}
\end{center}
\vspace{1cm}

\begin{center}{\Large\bf Abstract}
\end{center}
We consider the Wilson loop expectation in $SU(2)$ lattice gauge 
theory in the presence of constraints. 
The constraints eliminate from the functional measure gauge field 
configurations whose physical interpretation is that of 
thick center vortices linking with the loop. We give a simple proof that, 
for dimension $d\geq3$, the so constrained Wilson loop follows 
perimeter law, i.e. non-confining behavior, at weak coupling (low 
temperature). Thus the presence of vortex configurations is a 
necessary condition for confinement.

\vfill
\pagebreak  

\section{Introduction}
\setcounter{equation}{0}

The precise physical mechanism(s) by which $SU(N)$ gauge 
theories in their critical dimension apparently avoid a 
phase transition and thus remain in a confining phase for 
arbitrarily weak coupling is currently receiving renewed 
attention (see e.g. \cite{Lat97}). The proposal that extended 
thick vortices are the configurations responsible for this 
behavior has been strongly supported by recent numerical 
simulations in the case of $SU(2)$ \cite{KT}, \cite{Get}. 
(We refer to \cite{KT} for a recent discussion of the underlying 
heuristic physical picture, as well as earlier references to 
these ideas.) The string tension of the full Wilson loop at 
weak coupling is found in these simulations to be fully 
reproduced solely by the contribution of thick vortices. 
In this paper we give a simple proof of the converse statement, 
also seen in the simulations. We consider the 
Wilson loop in the presence of constraints eliminating field 
configurations which, in physical terms, represent thick vortices 
linking with the loop. We then show that the Wilson loop  
exhibits non-confining (perimeter law) behavior at weak coupling.

An earlier related result was obtained some time ago in \cite{Y}. 
There it is proven that, in the presence of constraints eliminating 
vortices winding around the (periodic) lattice, the 
electric-flux free energy order parameter exhibits nonconfining 
behavior at large $\beta$. Asymptotically large Wilson loops 
are indeed expected to be physically equivalent to the electric-flux 
order parameter. The latter, however, has been proven rigorously 
\cite{TY} only to form an upper bound on the Wilson loop. 
Thus it only provides a sufficient criterion for confinement: 
confining behavior of the electric-flux free energy implies 
confining behavior for the Wilson loop, but not the converse. 
At any rate, the electric-flux is an order parameter that 
refers to the entire lattice. It is clearly important to 
obtain a statement concerning the effect of the constraints 
on the Wilson loop itself, since it directly represents  the actual 
potential between two external sources (quarks), as the lattice 
is taken to the thermodynamic limit.  

The distinction between `thick', `thin' and `hybrid' 
vortices is crucial to the physical picture of the action of vortices 
at small gauge coupling (large $\beta$).
An appropriate framework for these considerations is obtained by 
expressing the $SU(N)$ lattice gauge theory in terms of new
separate $Z(N)$ and $SU(N)/Z(N)$ variables \cite{MP}, \cite{T1}. 
We review this formulation 
in section 2, where we also consider a variety of alternative 
lattice actions for which we obtain our result. We also 
introduce the appropriate constraints for eliminating linking thick 
and hybrid vortices. The proof that the so constrained Wilson 
loop follows perimeter law is given in section 3. We explicitly 
consider only the case of $SU(2)$ which exhibits all the  
relevant physical features of the general $N$ case.

{\it Notation.}\ We work on a simple hypercubic lattice 
$\Lambda \subset {\bf Z}^d$ of size $L_\mu = 2^{N_\mu}$, integer 
$N_\mu$, and lattice coordinates integer $n^\mu$, $\mu=1, \ldots, d$. 
Elementary cells $c^r$, $( r=0, 1, \ldots, d)$, in $\Lambda$ will 
be denoted more explicitly as $s$ (sites), 
$b$ (bonds), $p$ (plaquettes), $c$ (3-cubes), etc. Each $r$-cell 
is assigned an orientation; $-c^r$ then denotes the oppositely 
oriented cell. The number of cells in a collection $S$ of 
$r$-cells will be denoted by $|S|$. The oriented 
$r$-cells are the generators of the integer $r$-chains. There 
is a natural inner product on the cell complex defined by: 
$(c^r,c^q)=1$ if $r=q$, and $0$ otherwise. If then $\partial : 
c^r \to c^{r-1}$ is the 
boundary operator on $r$-chains, the coboundary operator 
$\partial^*$ is defined by: 
\[ (\partial^*c^r,c^{r+1}) = (c^r,\partial c^{r+1})\qquad. \]
Thus the coboundary of an $r$-cell 
$c^r$ is the $(r+1)$-chain formed of those $(r+1)$-cells (taken 
with appropriate orientation) whose boundaries contain $c^r$, 
i.e. $\pm c^{r+1} \in \partial^* c^r$  iff $\pm c^r\in\partial 
c^{r+1}$. By linearity this extends to arbitrary $r$-chains.

We employ the standard formalism of lattice gauge theory. 
With each bond we associate a copy $G_b$ of the gauge group $G$. 
The gauge field $U_b$ is an element of $G_b$, with the 
assignment $U_{-b} = U^{-1}_b$. The configuration space 
is then $\Omega_\Lambda=\otimes_{b\in\Lambda}G_b$, and a 
field configuration $U_\Lambda = \{U_b\}_{b\,\in\,\Lambda}$ 
is an element of $\Omega_\Lambda$. In the following we do not 
distinguish explicitly between the abstract group element 
$U_b$ and its matrix representation which, unless otherwise 
indicated, will always be the fundamental representation.

We will introduce various lattice variables taking values in 
the center of the group $G$.  
If $K$ is an abelian group, multiplicatively written, 
a  $K$-valued $r$-form $\alpha$ 
is the map: $\alpha: \;c^r\to\alpha[c^r]\in K$, 
with $\alpha[-c^r]= \alpha[c^r]^{-1}$. 
The exterior difference operator is then defined by 
\beq
(d\alpha)[c^{r+1}] = \prod_{c^r\in\partial c^{r+1}}\,
\alpha[c^r]\quad.
\eeq
Given any set $Q$ of $c^k$ cells, we generally employ the 
shorthand notation 
\beq
\alpha[Q]=\alpha[\prod_{k\in Q}c^k]=\prod_{k\in Q}\alpha[c^k]
\equiv \alpha_Q. 
\eeq 
Thus, e.g., for a 1-form $\gamma$ we write 
$\gamma_b, \;(d\gamma)_p$, and so on. In this paper $G=SU(2)$.

\section{Actions and constraints} 
\setcounter{equation}{0}
The Euclidean functional measure of the standard lattice $SU(2)$ 
theory is given by: 
\beq
d\mu_\Lambda(U) = Z_\Lambda^{-1}\:\prod_{b\in\Lambda}\,dU_b\;
\exp\left( \sum_p\,A_p(U_p)\right) \quad,\label{PF}
\eeq
where $dU_b$ denotes normalized Haar measure on $SU(2)$, 
and $A_p$ the plaquette action which is a function of 
$U_p=\prod_{b\in \partial p}U_b$, the product of bond variables $U_b$ 
around the plaquette $p$. The partition function $Z_\Lambda$ is 
defined by $\int d\mu_\Lambda(U) = 1$, and for any observable 
$F(U)$, i.e. (complex-valued) function on $\Omega_\Lambda$
\beq
\langle F(U) \rangle = \int\,d\mu_\Lambda(U)\,F(U)
\quad.\label{obs}
\eeq

The usual minimal plaquette action is the Wilson action 
\beq
A_p(U_p) = \beta\;\tr U_p \quad,\label{WA}
\eeq
which is a special case of the action 
\beq
A_p(U_p) = \beta\;\tr U_p + \lambda\;\mbox{sign}\,\tr U_p
\quad.\label{A1}
\eeq
(\ref{A1}) extrapolates between the Wilson action ($\lambda = 
0$) and the `positive plaquette model' action ($\lambda \to 
\infty$). Another action we consider in this paper is 
\beq
A_p(U_p) = \beta\;\tr U_p + \ln(\,\theta(\,|\tr U_p| - k\,)\,) 
\quad,\label{A2}
\eeq
where $\theta(x) = 1$ if $x>0$, 
$0$ if $x<0$, and $ 0 < k < 2$; in particular, $k$ a constant, 
or any function of $\beta$ such that $k\beta \to \infty$ as $\beta
\to \infty$, e.g. $k(\beta)=\beta^{-1/2}$.  
All these actions have, 
of course, the same naive continuum limit. More generally, 
they are all physically equivalent for sufficiently 
large $\beta$, where each plaquette becomes highly peaked 
around the perturbative vacuum $\tr U_p \to 1$. Use of such  
alternative actions provides a check of the requirement that 
long distance physics should not depend on short distance 
details such as the precise form of the latticized action.  
We will first prove our result working with the action (\ref{A2})
since, as it turns out, it allows a rather simpler proof of 
the result. We will then obtain the result for the more standard, 
but in fact physically equivalent, action (\ref{A1}).

To formulate our argument we rewrite the $SU(2)$ theory 
(\ref{PF}) in the $SO(3)\times Z(2)$ form \cite{MP}, \cite{T1}.  
Consider the configuration space $\Omega_\Lambda$ split
into equivalence classes, each class the coset bond variable 
configuration $\U_\Lambda=\{\U_b\}_{b\in\Lambda}$,  $\U_b\,\in\,
SU(2)/Z(2) \sim SO(3)$. Thus two configurations $U_\Lambda \ 
U^\prime_\Lambda \;\in \Omega_\Lambda$ are representatives of 
the same coset configuration $\U_\Lambda$ iff one has 
$U^\prime_b=U_b\gamma_b$, for some $\gamma \in Z(2), \ \forall 
b\in\Lambda$. Let   
$\eta_p(U)\, \equiv\, \mbox{sgn}\,\tr U_p $. Then 
\beq
(d\eta)_c(\U) = \prod_{p\,\in\,\partial c} \eta_p \label{deta}
\eeq
depends, as indicated, only on the coset configuration, since 
it is invariant under $U_b\to U_b\gamma_b$, $\gamma\in Z(2)$. 
Let $\sigma$ be a $Z(2)$-valued 2-form on $\Lambda$.
Adopt periodic boundary conditions (b.c.),  
and let $\{S^\alpha\}$, $\alpha=1, \ldots, {d \choose 2}$, be 
a set of 2-dimensional nonbounding closed surfaces forming a 
2-cycle basis on $\Lambda=T^d(N_1, \ldots, N_d)$. Then the 
$SU(2)$ theory (\ref{PF})-(\ref{obs}) 
can be expressed in the $SO(3)\times Z(2)$ form given by: 
\bea
d\mu_\Lambda(U,\sigma) & = & Z_\Lambda^{-1}\:\prod_{b\in\Lambda}
\,dU_b\prod_{p\in\Lambda}\,d\sigma_p\;\prod_{c\in\Lambda}\,
\delta\left(\,(d\eta)_c(d\sigma)_c\,\right)\, \prod_\alpha\, \delta\left(\,\eta[S^\alpha]\sigma[S^\alpha]\,\right)\nonumber\\
   &  & \quad\quad\quad\qquad\quad\quad\quad\quad 
\exp\left( \sum_{p\in\Lambda}
\,A_p(|{\rm tr}U_p|,\,\sigma_p)\right) \ .\label{altPF}
\eea 
In (\ref{altPF}) $\int\,d\sigma_p\,(\cdots) \equiv \frac{1}{2}
\sum_{\sigma_p=\pm1}\,(\cdots)$ denotes Haar measure over 
$Z(2)$, and
\beq 
\delta(\alpha) \equiv \int\,d\tau\,\chi_\tau(\alpha)= \frac{1}{2}
[\,1+\alpha\,]\quad, \label{Z2delta}
\eeq
defines the delta-function on the 
group $Z(2)$. Here $\chi_\tau(\alpha) =\alpha$ if $\tau=-1$, 
and $1$ otherwise, are the characters of $Z(2)$. The partition 
function $Z_\Lambda$ is defined by $\int d\mu_\Lambda(\U,
\sigma) = 1$. The plaquette action is given by 
\beq
A_p(|{\rm tr}U_p|,\,\sigma_p) = \beta\,|\tr U_p|\,\sigma_p 
+ \lambda\,\sigma_p \qquad\label{Aa}
\eeq
and 
\beq
A_p(|{\rm tr}U_p|,\,\sigma_p)  = \beta\;|\tr U_p|\,\sigma_p 
+ \ln(\,\theta\,(\,|\tr U_p| - k\,) \,)\label{Ab}
\eeq
corresponding to (\ref{A1}) and (\ref{A2}), respectively. 
(The result for the Wilson action is, of course, given by 
(\ref{Aa}) with $\lambda=0$.) (\ref{altPF}) is easily seen to be 
independent of the choice of basis $\{S^\alpha\}$. 
Note that (\ref{altPF}) is indeed a measure on the coset 
configuration space since it is invariant under $U_b \to 
U_b\gamma_b$, for any $Z(2)$-valued 1-form $\gamma$. 
If $\hat{F}(\U,\sigma)$ expresses a gauge invariant observable 
$F(U)$ in the new variables, the expectation (\ref{obs}) then 
satisfies 
\beq
\vev{F(U)} = \int\,d\mu_\Lambda(U,\sigma)\,\hat{F}(\U,\sigma) 
\eeq

Free b.c. on $U_{\partial\Lambda}$ results into (\ref{altPF}) 
without the product of $\delta$-functions over 2-cycles, and 
free b.c. on $\U_{\partial\Lambda}$, $\sigma_{\partial\Lambda}$. 
By selective omission of factors in this product, one may consider 
any mixture of free and periodic b.c. (\ref{altPF}) may be  
straightforwardly generalized to accommodate other useful 
types of b.c.,\footnote{In particular, `twisted' periodic b.c. 
on the torus given by action $\tau^\alpha 
A_p(U_p), \ \tau_\alpha\,\in\,Z(2)$, for $p\,\in\,S^\ast_\alpha$, 
where $\{S^\ast_\alpha\}$ a 2-cocycle basis, $(S^\ast_\alpha, 
S^\beta) = \delta_\alpha^\beta$, result into (\ref{altPF}) 
with periodic b.c. and the substitution $\delta(\tau_\alpha
\eta[S^\alpha]\sigma[S^\alpha])$ for each factor in the product 
over 2-cycles. Note that b.c. defined by summation over the twist
$\tau_\alpha$ eliminate the corresponding 
$\delta$-function factor. Such b.c. enter in the definition of 
the magnetic- and electric-flux free energy order parameters.
\label{Fbc}} and indeed applied to $\Lambda$ of any torsionless 
homology.

We are interested in the expectation of a Wilson loop 
subject to constraints whose physical effect is the 
elimination of its interaction with those 
vortices that are {\it not} energetically directly suppressed 
by the plaquette action at 
large $\beta$. These are extended vortices in the $\U_\Lambda$ 
configurations, `thick' vortices.\footnote{In the continuum 
extrapolation they are configurations topologically 
characterized by nontrivial element(s) of $\pi_1(SU(2)/Z(2)) 
= Z(2)$ ($\pi_1(SU(N)/Z(N))=Z(N)$ for general $N$).} Thin 
vortices, vortices in the $\sigma_\Lambda$ configurations, are 
directly suppressed by action cost proportional to their length.  
There are also hybrid vortices formed by the joining of 
`punctured' thick and thin vortices along their 
common coboundary which represents a magnetic current `loop', i.e.  
a  coclosed set of 3-cubes ${\cal L}$ with $(d\eta)_c=
(d\sigma)_c=-1,\ \forall c\in{\cal L}$. Such hybrids 
with a minimal short thin section but extended thick 
section can also affect long distance behaviour at 
large $\beta$. We refer to (\cite{KT}), and (\cite{Y}), 
(\cite{MP}), (\cite{T2}), for discussion of the physical 
interpretation of the constraints we now introduce.

We consider the constrained Wilson loop expectation:
\beq
W[C] = \Big<\: \tr U[C]\: \theta [\,\tr U[C]\,\eta_S\,] 
\:\Big> \label{WL}
\eeq
where now $\vev{-}$ denotes expectations in the restriction of the 
measure (\ref{PF}) to
\beq
d\mu_\Lambda(U) = Z_\Lambda^{-1}\:d\nu_\Lambda (U)\;
\exp\left( \sum_{p\in\Lambda}\,A_p(U_p)\right) \quad,\label{PF1}
\eeq 
with  
\beq
d\nu_\Lambda(U) = \prod_{b\in\Lambda}\,dU_b
\:\prod_{c\in\Lambda}\,\delta\,\left((d\eta)_c\right)\;\prod_\alpha\, \delta\left(\,\eta[S^\alpha]\,\right) \label{hnu}
\eeq 
and periodic b.c. ($Z_\Lambda$ defined by $\Big<\;1\;
\Big> = 1$). The constraint \cite{MP} 
\beq
\prod_c\;\delta\left(\,(d\eta)_c\right) \label{nomo}
\eeq
in the measure (\ref{PF1}) eliminates all 
monopoles, and hence all hybrid vortices in the theory. 
The factor 
\[  \theta [\,{\rm tr}U[C]\,\eta_S\,]\quad, \] 
where $\eta_S \equiv \prod_{p\,\in\,S}\:\eta_p$, which  
modifies the Wilson loop operator $\tr U[C]$ in (\ref{WL}), 
eliminates precisely all thick (unpunctured) vortices (while still 
allowing all thin vortices) linked with the loop \cite{Y}, \cite{KT}. 
Here $S$ is any surface bounded by the loop: $C=\partial S$. 
The expectation (\ref{WL}) is easily seen to be independent 
of the choice of $S$. The remaining constraint  
\[ \prod_\alpha\, \delta\left(\,\eta[S^\alpha]\,\right) \]
in (\ref{PF1}) is physically inessential,\footnote{It is in fact 
automatically enforced as $|\Lambda|\to\infty$ at large $\beta$ 
because of the energetic suppression of thin vortices  
winding completely around the lattice. This can be seen clearly from 
(\ref{WL0}) below.} and only included for 
technical convenience in adopting periodic b.c. It can be 
eliminated by adopting free b.c. or twisted plus 
untwisted periodic b.c. (cp. footnote \ref{Fbc}).

We pass then to $SO(3)\times Z(2)$ variables in which (\ref{WL}) 
assumes the form 
\bea
W[C] & = & Z_\Lambda^{-1}\,\int\,d\nu_\Lambda(U)\prod_{p\in
\Lambda}\,d\sigma_p\, \prod_{c\in\Lambda} \delta\left((d\eta)_c
(d\sigma)_c\right)\; \prod_\alpha\, \delta\left(\,\eta[S^\alpha]\sigma[S^\alpha]\,\right)\nonumber\\
    &  &  \qquad \quad\cdot\, \exp\left( \sum_p\,
A_p(|\tr U_p|,\,\sigma_p)\right)\;\,\tr U[C]\,\eta_S\; \sigma_S \;
\theta[\,\tr U[C]\,\eta_S\,] \label{WL0}\\
   & = &  Z_\Lambda^{-1}\,\int\,d\nu_\Lambda(U)\prod_{b\in\Lambda}
\,d\gamma_b\, \exp\left( \sum_p\,A_p(\,
|\tr U_p|,\,(d\gamma)_p\,)\right) \nonumber\\
  &  & \qquad \qquad \qquad\qquad  \theta[\,{\rm tr}U[C]\,\eta_S\,]
\; \tr U[C]\,\eta_S \,\gamma[C]\, \quad.\label{WL1}
\eea
where $\gamma$ a $Z(2)$-valued 1-form. 
The second equality is obtained by solving the constraints on 
$\sigma_\Lambda$ that result from the restriction to (\ref{hnu}): 
\bea
(d\sigma)_c=1\ , \ \forall c \,,\quad \delta\big(\sigma[S^\alpha]\big)
 = 1\,,\ \forall \alpha \ &\Longrightarrow &\ \sigma_p = 
\prod_{b\in \partial p}\,\gamma_b = (d\gamma)_p \nonumber\\ 
  &  & \sigma_S\equiv \prod_{p\,\in\,S}\sigma_p = \prod_{b\in C}
\gamma_b = \gamma[C]\;.
\eea 
Now, 
\beq 
\theta[\,\tr U[C]\,\eta_S\,] \,\tr U[C]\,\eta_S  = 
\theta[\,\tr U[C]\,\eta_S\,] \,|\tr U[C]|\quad, 
\eeq
so we can write (\ref{WL}) in the form 
\beq
W[C] = \Big\langle\; \theta[\,{\rm tr}U[C]\,\eta_S\,] 
|\tr U[C]|\,\gamma[C] \;\Big\rangle_{SO(3)\otimes Z(2)}
\quad,\label{WL2}
\eeq
where 
\beq
\Big\langle\; - \;\Big\rangle_{SO(3)\otimes Z(2)} = 
\int\,d\mu_\Lambda(U,\gamma)
\; (-) \label{ex1}
\eeq 
with 
\beq
d\mu_\Lambda(U,\gamma)  \equiv  Z_\Lambda^{-1}
\;d\hnu_\Lambda(U)\;\prod_{b\in\Lambda}\,d\gamma_b\;
\exp\left( \sum_{p\,\in\,\Lambda}\,K_p(U)(d\gamma)_p\,\right) 
\label{mes}
\eeq
and periodic b.c. $d\hnu_\Lambda(U)$ and $K_p(U)=K_p(\U)$ are 
defined as
\bea
d\hnu_\Lambda(U) & \equiv & d\nu_\Lambda(U)\nonumber\\
K_p(U) & \equiv & \beta\,|\tr U_p| + \lambda \label{mesAa}
\eea
for the action (\ref{Aa}), and 
\bea
d\hnu_\Lambda(U) & \equiv & d\nu_\Lambda(U)\,\prod_{p\in\Lambda}
\theta\,(\,|\tr U_p| - k\,)\nonumber\\
K_p(U) & \equiv & \beta\,|\tr U_p| \label{mesAb}
\eea
for the action (\ref{Ab}). It is easily verified that  
(\ref{mes}) is a reflection positive measure in $(d-1)$-dimensional 
planes with sites. 

Our task in the following is to bound the constrained 
Wilson loop expectation (\ref{WL2}) from below.

\section{Absence of confinement} 
\setcounter{equation}{0}

We formulate our main result as
\begin{theorem} For sufficiently large $\beta$, and dimension 
$d\geq 3$ the constrained Wilson loop expectation (\ref{WL}), 
or, equivalently, (\ref{WL2}), exhibits perimeter law, i.e. 
there exist constants $\alpha,\ \alpha_1(d),\ \alpha_2(d)$ 
such that  
\beq 
W[C] \,\geq \, \alpha \exp\left(\, - \alpha_2\,e^{-\alpha_1\beta}
\,|C|\,\right) \quad.\label{len}
\eeq 
\label{Thm}
\end{theorem}
From now on all expectation signs are meant in the measure 
(\ref{mes}), and, for brevity, we drop the label 
$SO(3)\otimes Z(2)$.  

Choose the loop $C$ so that it is bisected into two equal pieces 
by a $(d-1)$-dim plane with sites. Then reflection positivity (RP) 
of the measure (\ref{mes}) implies
\beq 
\Big\langle\;\tr U[C]\,\eta_S\:\gamma[C]\;\Big\rangle \geq 0
\quad.\label{pos}
\eeq
(Without loss of generality, the surface $S$ may always be 
assumed to be also reflection symmetric in the hyperplane 
bisecting $C$.) Inserting 
\[ 1 = \theta[\,\tr U[C]\,\eta_S\,] + 
\theta[\,-\tr U[C]\,\eta_S\,] \]
in (\ref{pos}) we then have 
\bea
0 & \leq &  \Big\langle\;\theta[\,\tr U[C]\,\eta_S\,]
\:\tr U[C]\: \eta_S \:\gamma[C]\;\Big\rangle + \Big\langle\;
\theta[\,- \tr U[C]\,\eta_S\,]\:\tr U[C]\,\eta_S\,\gamma[C]
\;\Big\rangle \nonumber\\
   & = &  \Big\langle\;\theta[\,\tr U[C]\,\eta_S\,]\:|\tr U[C]|
\,\gamma[C]\;\Big\rangle - \Big\langle\;\theta[\,- 
\tr U[C]\,\eta_S\,]\:|\tr U[C]|\,\gamma[C]\;\Big\rangle 
\quad.\label{pos1}
\eea
On the other hand 
\beq
\Big\langle\;|\tr U[C]|\,\gamma[C]\;\Big\rangle = 
\Big\langle\;\theta[\,\tr U[C]\,\eta_S\,]\:|\tr U[C]|
\,\gamma[C]\;\Big\rangle + \Big\langle\;
\theta[\,- \tr U[C]\,\eta_S\,]\:|\tr U[C]|\,\gamma[C]\;
\Big\rangle. \label{dec}
\eeq
Adding (\ref{pos1}) and (\ref{dec}) we have 
\beq
\Big\langle\;\theta[\,\tr U[C]\,\eta_S\,]\:|\tr U[C]|\,\gamma[C]
\;\Big\rangle  \geq \frac{1}{2}\;\Big\langle\;|\tr U[C]|\,
\gamma[C]\;\Big\rangle \quad.\label{ineq1}
\eeq
Hence from (\ref{WL2}) we obtain:  
\beq
W[C] \geq \frac{1}{2}\;\Big\langle\;|\tr U[C]|\,
\gamma[C]\;\Big\rangle \quad.\label{WL3}
\eeq
Now 
\bea
\frac{1}{2}\;\Big\langle\;|\tr U[C]|\,
\gamma[C]\;\Big\rangle 
 & = & Z_\Lambda^{-1}\,\int d\hnu_\Lambda(U)\,\frac{\D 
|\tr U[C]|}{\D 2}\nonumber\\
   & & \qquad\quad\cdot\int\,\prod_{b\in\Lambda}\,d\gamma_b\:
\gamma[C]\,\exp\left(\sum_{p\in\Lambda}K_p(U)(d\gamma)_p\right) 
\nonumber\\
    & = & \int\, d\hmu_\Lambda(U)\,\frac{\D
|\tr U[C]|}{\D2}\;\vev{\,\gamma[C]\,}_{\ssc Z(2)}(\{K_p(U)\}) 
\;,\label{Z2ex1}
\eea
where 
\beq
d\hmu_\Lambda(U) \equiv Z^{-1}_\Lambda\;d\hnu(U) Z_{\ssc Z(2),
\Lambda}(\{K_p(U)\})  \quad,\label{hmes}
\eeq 
and 
\beq
\Big\langle\;-\;\Big\rangle_{\ssc Z(2)}(\{K_p\}) = 
Z_{\ssc Z(2),\Lambda}^{-1}
(\{K_p\})\,\int\,\prod_{b\in\Lambda}d\gamma_b\,(\;-\;)\, 
\exp(\sum_{p\in\Lambda}K_p\gamma_p) \label{Z2ex2}
\eeq
denotes expectations in the $Z(2)$ lattice gauge theory 
with plaquette couplings $K_p$, and partition function 
$Z_{\ssc Z(2),\Lambda}$ defined  by $\vev{\;1\;}_{\ssc Z(2)}
(\{K_p\})=1$. In (\ref{Z2ex1}) we have the 
$U$-dependent couplings $K_p(U) \geq 0$ given by (\ref{mesAa}), 
(\ref{mesAb}). By the Griffiths inequalities \cite{Gr} 
applied to (\ref{Z2ex2}): 
\beq
\vev{\,\gamma[C]\,}_{\ssc Z(2)}(\{K_p(U)\})\geq 0 
\quad,\label{G1}
\eeq 
whereas
\beq
\frac{1}{2}|\tr U[C]| \geq \frac{1}{4}|\tr U[C]|^2 = 
\frac{1}{4}|\chi_F(U[C])|^2 = \frac{1}{4}
\big[\;\chi_A(U[C]) + 1\;\big] \label{chi}
\eeq 
where $\chi_F$ and $\chi_A$ denote the $SU(2)$ characters in 
the fundamental and adjoint representation, respectively. 
Combining (\ref{G1}), (\ref{chi}) with (\ref{Z2ex1}) then gives 
\bea
\frac{1}{2}\;\Big\langle\;|\tr U[C]|\,
\gamma[C]\;\Big\rangle 
 & \geq & \frac{1}{4} \int\, d\hmu_\Lambda(U)\;
\vev{\,\gamma[C]\,}_{\ssc Z(2)}(\{K_p(U)\}) \nonumber\\
  & & \quad\quad\quad + \frac{1}{4} \int\, d\mu_\Lambda(U,\gamma)
\;\chi_A(\,U[C]\,)\;\gamma[C]\quad.\label{ineq2} 
\eea 
Now by RP the 2nd term on the r.h.s. in (\ref{ineq2}) is 
non-negative. So, from (\ref{WL3}), we obtain: 
\beq
W[C] \geq \frac{1}{4} \int\, d\hmu_\Lambda(U)\;
\vev{\,\gamma[C]\,}_{\ssc Z(2)}(\{K_p(U)\})
\quad. \label{WL4}
\eeq 

{\it Action (\ref{Ab})}.\ Completing the proof of the result 
in the case of the action (\ref{Ab}) is now straightforward. 
Applying again Griffiths inequalities, one has
\beq
\vev{\,\gamma[C]\,}_{\ssc Z(2)}(\{K_p\}) \geq 
\vev{\,\gamma[C]\,}_{\ssc Z(2)}(\{K_p^\prime\})\ , 
\qquad\; K_p \geq K_p^\prime\;, \forall\; p\quad.\label{comp}
\eeq
With the measure given by (\ref{mes}), (\ref{mesAb}), (\ref{hmes}), 
use of (\ref{comp}) in (\ref{WL4}) gives 
\beq
W[C] \geq \frac{1}{4} \;\vev{\,\gamma[C]\,}_{\ssc Z(2)}(k\beta) 
\; . \label{WLineq}
\eeq
The r.h.s. is the loop expectation in the $Z(2)$ gauge theory with  
$K_p=k\beta,\;\, \forall\:p$.  

It is a well-known result that, for sufficiently large $\beta$ 
and $d\geq3$: 
\beq
\vev{\,\gamma[C]\,}_{\ssc Z(2)}(\beta) \geq 
\mbox{Const.} \;\exp(-\rho(\beta)\,|C|) 
\;,\label{Z2len}
\eeq
i.e. the expectation exhibits perimeter-law. A rigorous 
proof (see e.g. Thm. 3.17 in \cite{Sei}, \cite{M-MS}) is
by standard polymer expansion either as low-temperature 
expansion in `contours',\footnote{A special case of 
the expansion we use in treating the action (\ref{Aa}) below.}  
or, after a duality transformation, as high-temperature expansion. 
Hence, (\ref{WLineq}) gives perimeter-law lower bound on $W[C]$ 
for all $\beta$ such that $k\beta$ large enough. 
This conclude the proof of the theorem in the case of the 
action (\ref{Ab}).

{\it Action (\ref{Aa})}. A coclosed set of plaquettes which 
cannot be decomposed into two disjoint coclosed sets will be 
called a contour. (Two plaquettes are defined to be disjoint if 
they share no link.) Given a configuration $\{\gamma_b
\}_{b\,\in\,\Lambda}$, the set of plaquettes on which 
$(d\gamma)_p = -1$ is a coclosed set ($(d-2)$-dim closed set 
on the dual lattice $\Lambda^\ast$). Such a coclosed set 
$\Gamma$ can be uniquely decomposed into a disjoint 
contours $\zeta_1, \ldots, \zeta_{|\Gamma|}$. Each such contour 
then is the site of a thin vortex. We expand the expectation $\vev{\,\gamma[C]\,}_{\ssc Z(2)}(\{K_p(U)\})$ in (\ref{WL4}) 
in a contour expansion. The partition function in the 
denominator is expanded in the form: 
\beq
Z_{\ssc Z(2),\Lambda}(\{K_p(U)\}) = \left(2^{-|\Lambda|} 
\prod_{p\,\in\,\Lambda}e^{K_p(U)}\right)\; \sum_{\Gamma \,
\subset\,\Lambda}\,z_\Gamma(U) \;. \label{CPF}
\eeq 
The sum is over all sets $\Gamma$ of disjoint 
(compatible) contours $\{\zeta_1, \ldots, \zeta_{|\Gamma|}\}$ 
and
\beq
z_\Gamma(U) \equiv \prod_{\zeta\,\in\,\Gamma}\,z_\zeta(U)
\quad,\label{Gactiv}
\eeq
with
\beq
z_\zeta(U) \equiv \prod_{p\,\in\,\zeta} \exp\big(\,-2K_p(U)\,
\big) \label{Cactiv}
\eeq
the activity of contour $\zeta$, and $K_p(U)$ given by 
(\ref{mesAa}). Noting that 
$\vev{\,\gamma[C]\,}_{\ssc Z(2)}(\{K_p(U)\})= 
\vev{\,\gamma[S]\,}_{\ssc Z(2)}(\{K_p(U)\})$, 
where $\partial S = C$, it is easily seen that the expansion 
of the numerator
\[ Z^C_{\ssc Z(2),\Lambda}(U) \equiv Z_{\ssc Z(2),\Lambda}(U)\; \vev{\,\gamma[C]\,}_{\ssc Z(2)}(\{K_p(U)\}) \] is 
given by the expansion (\ref{CPF}) after replacement of the 
contour activities $z_\zeta(U)$, eq. (\ref{Cactiv}), by 
\beq
z^C_\zeta(U) = (-1)^{l(C,\zeta)}\,z_\zeta(U)\quad,\qquad l(C,
\zeta) = |S\cap\zeta| \quad\ (\mbox{mod}\ 2)\quad.\label{CCactiv}
\eeq 
(Here $l(C,\zeta)$ is the (mod 2) linking number of the contour 
$\zeta$ with the loop $C$.) 

Letting 
\beq
\Z_{\ssc Z(2),\Lambda} = \sum_{\Gamma \,
\subset\,\Lambda}\,z_\Gamma(U)\;\;,\qquad \Z^C_{\ssc Z(2),\Lambda}
= \sum_{\Gamma \,\subset\,\Lambda}\,z^C_\Gamma(U)\;\ ,\label{clsums}
\eeq
and applying Jensen's inequality in (\ref{WL4}) gives 
\bea 
W[C] & \geq & \frac{1}{4} \int\, d\hmu_\Lambda(U)\;
\exp \Big( \ln \Z^C_{\ssc Z(2),\Lambda}(U) - 
\ln \Z_{\ssc Z(2),\Lambda}(U)\Big)
\nonumber\\ 
 & \geq & \frac{1}{4}\exp \left( \int\, d\hmu_\Lambda(U)\;
\Big( \ln \Z^C_{\ssc Z(2),\Lambda}(U) - \ln \Z_{\ssc Z(2),
\Lambda}(U)\Big)\right)\nonumber \\
       & \equiv & \frac{1}{4}\exp {\cal F}(\beta,\lambda)\quad.
\label{WL5} 
\eea 
Noting that $\Z_{\ssc Z(2),\Lambda}(U) =1$ at $z_\zeta(U)=0$, 
we may define $\ln \Z_{\ssc Z(2),\Lambda}(U)$ as that 
continuous branch of the logarithm which vanishes at 
vanishing contour activity, and similarly for 
$\ln \Z^C_{\ssc Z(2),\Lambda}(U)$. A closed form 
representation of ${\cal F}(\beta,\lambda)$ is given by the 
M\"{o}bius inversion representation of 
$\ln \Z_{\ssc Z(2),\Lambda}(U)$  and of 
$\ln \Z^C_{\ssc Z(2),\Lambda}(U)$ as a finite series (for 
finite $\Lambda$) of logarithms of partition function of 
linked clusters of single multiplicity \cite{KP}. 
Expanding these logarithms leads to the standard 
expansion of linked clusters of repeated multiplicities: 
\beq 
{\cal F}(\beta,\lambda) = \int\, d\hmu_\Lambda(U)
\;\left(\:\sum_{Q\,\subset\,\Lambda}\:a(Q)\,
\Big(\,z^C_{Q}(U) - z_{Q}(U)\,\Big)\:\right)\quad .\label{F1}
\eeq 
The sum is over all linked clusters $Q=\{\zeta_1,\ldots,
\zeta_{n_{Q}}\}$ of contours with the property that at least 
one contour in each $Q$ winds around $C$. Multiple copies of 
a contour are allowed to appear as distinct members in a cluster, 
and 
\beq 
 a(Q) = \sum_{G(Q)} (-1)^{|G(Q)|} \quad.\label{comb}
\eeq 
Here the sum is over all connected graphs $G(Q)$ on the 
vertex set $\{\zeta, \ldots, \zeta_{n_{Q}}\}$ with a line 
connecting two vertices if they represent incompatible (not 
disjoint) contours. $|G(Q)|$ denotes the number of lines 
in $G(Q)$.
   
The expansion in (\ref{F1}) converges for sufficiently small 
activities  $z_\zeta(U)$. Let $|z_\zeta(U)| \leq \exp( - b\,|
\zeta|)$. The number of contours of size $q$ with one plaquette 
fixed is bounded by $[10(d-2)]^{10(d-2)q}$. Applying well-known 
results on the convergence of the polymer-type cluster expansion  
\cite{B}, \cite{Cam}, it follows that the expansion on the 
r.h.s. of (\ref{F1}) is absolutely convergent, uniformly in 
$|\Lambda|$ and over $\Omega_\Lambda$, provided 
\beq 
b \geq 10(d-2)\ln 10(d-2) + \ln 5 \quad;\label{conv}
\eeq
hence, by (\ref{mesAa}), $\lambda$ sufficiently large. Uniform 
convergence allows integration term by term, so we have  
\beq 
{\cal F}(\beta,\lambda) = \sum_{Q\,\subset\,\Lambda}\:a(Q)\,
\Big(\,\vev{ z^C_{Q}(U)} - \vev{ z_{Q}(U)}\,\Big)
\quad .\label{F2}
\eeq  

We now observe that, for $\beta$ sufficiently large, the 
series (\ref{F2}) converges for all $\lambda \geq 0$. 
To show this, we bound $\vev{z_Q(U)}$ uniformly in $|\Lambda|$ 
by chessboard estimates \cite{Fr}. One finds 
(Appendix)
\beq
\vev{z_Q(U)} \ \leq \ \prod_{\zeta\,\in\,Q}\,
\hat{z}(\beta,\lambda)^{|\zeta|} \label{chess}
\eeq 
where 
\beq
\hat{z}(\beta,\lambda) = c_1e^{-c_2\beta - 2\lambda}
\eeq 
with positive constants $c_1(d),\ c_2(d)$ depending only
on the dimension $d$. It follows that (\ref{F2}) converges 
absolutely and uniformly in $|\Lambda|$, provided 
$|\hat{z}(\beta,\lambda)| \leq e^{-b}$ with $b$ satisfying   
(\ref{conv}); hence  for $\beta$ large enough, and all $\lambda
\geq0$. Uniqueness of analytic continuation 
then implies that the representation of  ${\cal F}$ 
by the cluster expansion in 
(\ref{F2}), originaly obtained in the domain of large 
${\rm Re}\ \lambda$, is valid in this   
extented convergence domain ${\rm Re}\ \lambda \geq 0$. 

The leading contribution to (\ref{F2}) comes from the 
shortest possible contours, each consisting of $2(d-1)$ 
plaquettes forming the coboundary of a bond on the loop $C$, 
and there are $|C|$ of them. A bound on the remainder   
by the same as leading ${\cal O}(|C|)$-type behavior is a 
standard corollary of the polymer expansion convergence 
proof (e.g. \cite{Sei}). This concludes the proof of the 
Theorem \ref{Thm} for the action (\ref{Aa}).

\setcounter{section}{0}     
\renewcommand{\thesection}{\Alph{section}}
\renewcommand{\theequation}{\mbox{\thesection.\arabic{equation}}}
\section{Appendix}
\setcounter{equation}{0}
 
Let $\Pi_e$ denote the set of `even' 2-dimensional $[
\kappa\lambda]$-planes: $x^\mu = 2n^\mu$, integer $n^\mu$, 
$\mu\neq \kappa,\ \lambda$. If $|\Pi_e|$ and $|\Lambda|$ denote 
the number of plaquettes in $\Pi_e$ and $\Lambda$, respectively, 
we have $|\Pi_e| = |\Lambda|/s$, where $s = 2^{d-3}d(d-1)$. 
Consider now a cluster $Q$, and let $\zeta_e \equiv \zeta 
\cap \Pi_e$. Using RP to reflect repeatedly in 
$(d-1)$-dimensional planes with sites $x^\mu = (2n^\mu+1)$, 
\ $\mu\neq\kappa,\lambda$;\ $x^\mu = n^\mu, \ \mu=\kappa, 
\lambda$, and the fact that $\exp (-2\beta|\tr U_p|) \leq 1$, 
one straightforwarly obtains the chessboard estimate 
\bea 
\vev{z_Q(U)} & = & e^{-2\lambda|Q|} \vev{\,\prod_{\zeta\,
\in\,Q}\prod_{p\,\in\,\zeta}\,e^{-2\beta|\tr U_p|} \,} 
\nonumber\\
   & \leq & e^{-2\lambda|Q|} \vev{\,\prod_{\zeta\,
\in\,Q}\prod_{p\,\in\,\zeta_e}\,e^{-2\beta|\tr U_p|} \,} 
\nonumber\\   
  & \leq & e^{-2\lambda|Q|} \vev{\,\prod_{p\,\in\,\Pi_e}
\,e^{-2\beta|\tr U_p|} \,}^{|Q_e|/|\Pi_e|} \label{chess1}\quad, 
\nonumber\\
\eea 
where $|Q| = \sum_{\zeta\,\in\,Q} |\zeta|$, and $|Q_e| = 
\sum_{\zeta\,\in\,Q} |\zeta_e|$. 
To estimate the last expectation on the r.h.s. in (\ref{chess1})  
it now suffices to bound the numerator from above by 
its maximum :  
\beq 
\vev{\,\prod_{p\,\in\,\Pi_e}\,e^{-2\beta|\tr U_p|} \,} 
\leq Z_\Lambda^{-1} \, e^{\lambda|\Lambda|} e^{2\beta\,( 
|\Lambda| - |\Pi_e|)} \quad.\label{num} 
\eeq 
The partition function in the denominator is bounded 
from below by restricting $\gamma_b$ to $1$, 
and each $U_b$ integration to 
a small neighborhood of the identity such that $|\tr U_p| \geq 
2e^{-\delta}$ for all $p$. Let $\tau_\delta$ be the volume of 
this neighborhood. The constraints in $d\hnu_\Lambda(U)$ 
are then automatically satisfied, and we have 
\beq 
Z_\Lambda \geq \left(\left({\tau_\delta\over 2}\right)^{2/(d-1)} \,
e^{2\beta e^{-\delta}+ \lambda}\right)^{|\Lambda|}
\; .\label{denm}  
\eeq
Now, given a cluster $Q$, we are free to pick the 
definition\footnote{We are free to choose `even' coordinates 
(choice of lattice coordinate origin) and directions 
$[\kappa\lambda]$.} of the `even' planes $\Pi_e$ so that  
$|\zeta_e| \geq |\zeta|/s$, i.e $|Q_e|\geq |Q|/s = |Q|/ 
2^{(d-3)}d(d-1)$. Combining this with (\ref{num}), (\ref{denm}), 
we find 
\beq
\vev{\,\prod_{p\,\in\,\Pi_e}\,e^{-2\beta|\tr U_p|} \,} ^{
|Q_e|/|\Pi_e|}\ \leq\  \hat{a}(\beta)^{|Q|} \label{chess2}\quad,
\eeq
where we defined 
\beq 
\hat{a} \equiv \min_\delta \left(\,\left({2\over\tau_\delta}\right)
^{2/(d-1)} \,\exp\Big(-{2\beta\over s}[\:1 - s(1-e^{-\delta})\:]
\Big)\right) 
\label{hata}\quad.
\eeq
Thus $\hat{a}(\beta) \to 0$ exponentially as $\beta\to\infty$. 
(\ref{chess1}) with (\ref{chess2}) and
\beq 
\hat{z}(\beta,\lambda)= e^{-2\lambda}\,\hat{a}(\beta) 
\eeq 
then produces the bound (\ref{chess}). 
 
A more refined chessboard estimate, utilizing reflections onto 
every $d$-cell (dual site) of the lattice is possible (cp 
\cite{Y}), and results into elimination of the factor $1/s$ 
multiplying $\beta$ in the exponent in (\ref{hata}), a 
substantial numerical improvement. The simpler estimate given 
here, however, suffices for our purposes. 

\pagebreak

\end{document}